# Spin magnetic susceptibility in the two-layer Hubbard model

Ferdinando Mancini [a,b], Volodymyr Turkowski [a,b,1]

[a] *Dipartimento di Scienze Fisiche "E.R.Caianiello" e Unitá I.N.F.M. di Salerno Universitá di Salerno, 84081 Baronissi (SA), Italy*

[b] *IIASS "E.R.Caianiello", 84019 Vietri sul Mare (SA), Italy*

**Abstract**

The two-layer Hubbard model is studied by using the composite operator method. Magnetic properties of the model in the normal state are analyzed by studying the uniform static spin magnetic susceptibility as a function of doping and temperature. The relevance of the model to describe the physical behavior of two-layered cuprate high-$T_c$ superconductors is discussed.

## 1. Introduction

The theoretical description of high-$T_c$ superconductors (HTSC) still remains one of the most difficult and important tasks of the modern solid-state physics. There is a widespread opinion that antiferromagnetic correlations in the $Cu - O$ planes play an important role in the behavior of these systems. In this sense it is interesting to study the magnetic properties of HTSC in the framework of some strongly correlated electronic model. It was shown (see, for example, Refs.[1,2]), that the magnetic normal properties of the lanthanium family materials can be qualitatively described by using the $2D$ Hubbard model in the composite operator framework.

The aim of this paper is to study such properties, namely the uniform static spin magnetic susceptibility, in the case of the two-plane Hubbard model which, we believe, can describe the behavior of the two-layered yttrium family materials.

## 2. Results

The two-plane Hubbard model is described by the next Hamiltonian

$$H = \sum_{i,j,\sigma}(t_{ij} - \mu\delta_{ij})c_\sigma^\dagger(i)c_\sigma(j) + U\sum_i n_\uparrow(i)n_\downarrow(i).$$

Here we use the standard notations; the nearest hopping constant takes two values $t$ and $t_z$ for intra- and inter- plane hopping, correspondingly.

The composite operator method calculations give the next expression for the spin magnetic susceptibility in the static approximation

$$\chi(\mathbf{k},\omega) = \frac{2}{n^2 - n - 2D}[n(Q_{1111}^R + 2Q_{1112}^R + Q_{1212}^R) + (2-n)(Q_{1212}^R + 2Q_{1222}^R + Q_{2222}^R)],$$

[1] Corresponding author. E-mail: vturk@sa.infn.it

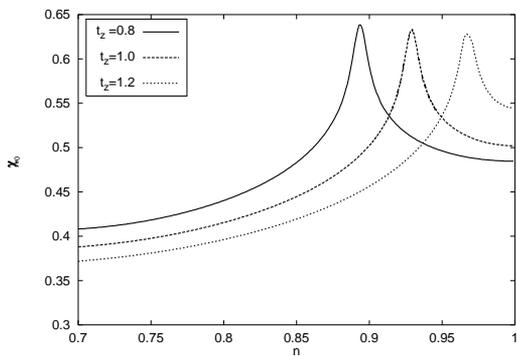

Fig. 1. The values of $\chi_0$ are plotted as a function of $n$ for different values of $t_z$ at $U/t = 4$ and $T/t = 0.01$.

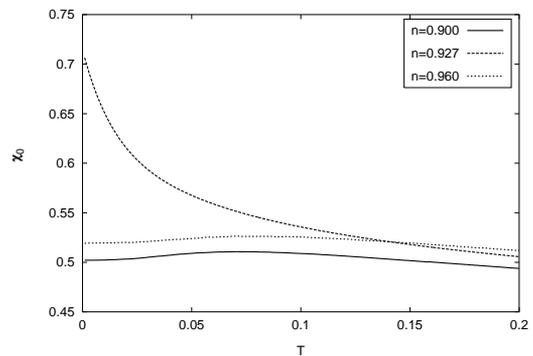

Fig. 2. The values of $\chi_0$ are plotted as a function of $T$ for different values of $n$ at $U/t = 4$ and $t_z = 1$.

where $n$ is the particle density and $D$ is the double occupancy.

The function $Q^R_{\alpha\beta\gamma\delta}(\mathbf{k},\omega)$ is the retarded part of the Fourier transformed expression of $Q_{\alpha\beta\gamma\delta}(i,j) = G_{\alpha\beta}(i,j)G_{\gamma\delta}(j,i)$. The thermal causal Green's functions (GFs) $G(i,j)$ are defined by $G(i,j) = <T[\Psi(i)\Psi^\dagger(j)]>$ in the fermionic basis $\Psi^\dagger(i) = (\xi^\dagger(i), \eta^\dagger(i))$, where $\xi(i) = [1-n(i)]c(i)$ and $\eta(i) = n(i)c(i)$ are the Hubbard's operators.

As it was shown previously (e.g. [3]), the single particle retarded GFs can be determined self-consistently with requirement to satisfy the Pauli principle $\langle\xi(i)\eta^\dagger(i)\rangle = 0$.

In this paper we present the calculations of the uniform static spin magnetic susceptibility $\chi_0 = \chi(\mathbf{k} \to 0, \omega = 0)$. In Fig.1 the values of $\chi_0$ are plotted versus $n$ for different values of the hopping constant $t_z$. Susceptibility $\chi_0$ demonstrates a maximum at some value of $n_c$ which depends on $t_z$. As in the 2D case, this peak is a consequence of the van Hove singularity (vHs) in the density of states. Numerical studies show that the value of $n_c$ increases almost linearly with $t_z$ and $n_c = 1$ at $t_z \simeq 1.4$ in the case $U/t = 4$ and $T/t = 0.01$. The experimental results [4] also demonstrate the maximum for $\chi_0(n)$ in $YBa_2Cu_3O_{7-x}$.

In Fig.2 $\chi_0$ as a function of temperature is presented for different values of doping. Qualitatively the same behavior for $\chi_0$ was found experimentally for some double-layered materials [4,5]. As it also follows from this picture, $\chi_0$ demonstrate a sharp maximum at $T = 0$ when $n = n_c$.

It should be mentioned that the position of the vHs also changes with $t'$ in the two-dimensional $t - t' - U$ Hubbard model [6,7]. However, the composite operator studies show [8] that for obtaining the right critical value $n_c$, in accordance with experiments on ittrium family materials, $t'$ should be taken positive $\sim 0.5$, whereas it is impossible with positive values to reproduce the right form of the Fermi surface after the ARPES data [9]. In this sence, from the composite operator method point of view, the two-layer Hubbard model looks more promising for describing the normal properties of the ittrium family materials.